\documentclass[10pt]{article}

\usepackage{graphicx}
\usepackage{amssymb,amsmath,amsfonts}

\newcommand{\f}[2]{{\ensuremath{\mathchoice
{\dfrac{#1}{#2}}
{\dfrac{#1}{#2}}
{\frac{#1}{#2}}
{\frac{#1}{#2}}
}}}

\usepackage{opex3}
\begin{document}

\title {{\it M}-lines characterization of selenide and telluride waveguides for mid-infrared interferometry}

\author{Lucas Labadie}
\address{Max-Planck Institut f\"ur Astronomie, K\"onigstuhl 17, D-69117 Heidelberg, Germany}
\email{labadie@mpia.de}
\address{Laboratoire d'Astrophysique de l'Observatoire de Grenoble, BP 53, 38041 Grenoble C\'edex, France}

\author{Caroline Vigreux-Bercovici and Annie Pradel}
\address{Laboratoire de Physico-Chimie de la Mati\`ere Condens\'ee, Institut
  Gehrardt, UMR 5617, Place Eug\`ene Bataillon, 34095 Montpellier C\'edex 5,
  France}

\author{Pierre Kern and Brahim Arezki}
\address{Laboratoire d'Astrophysique de l'Observatoire de Grenoble, BP 53,
  38041 Grenoble C\'edex, France}

\author{Jean-Emmanuel Broquin}
\address{Institut de Micro-\'electronique et Photonique, UMR 5130, BP 257,
  38016 Grenoble C\'edex 1, France}
\vspace{0.25cm}
\begin{abstract} Nulling interferometry is an astronomical technique that combines equal wavefronts to achieve a deep rejection ratio of an on-axis star, and that could permit to detect Earth-like planets in the mid-infrared band [5 -- 20 $\mu$m]. Similarly to what is done in the near-infrared, high frequencies spatial filtering of the incoming beams can be achieved using single-mode waveguides
  operating in the mid-infrared. An appreciable reduction of the instrumental
  complexity is also possible using integrated optics (IO) devices in this
  spectral range. The relative lack of single-mode guided optics in the mid-infrared
  has motivated the present technological study to demonstrate the feasibility
  of dielectric waveguides functioning at longer wavelengths. We propose to
  use selenide and telluride components to pursue the development of more
  complex IO functions.
\end{abstract}




%






%


\section{Introduction}

In the last ten years, the photonics field has provided a number of valuable technological
solutions to astronomers to improve the performance of the observatories
instruments. In particular, optical fibers and integrated optics (IO) devices have
already been part \cite{Foresto, Berger}, or will be part in the future
\cite{Malbet, Perrin} of interferometric facilities. As a result, this context has
raised the question of extending the use of photonics devices to the Darwin
Infrared Space Interferometer of the {\it European Space Agency} (ESA) \cite{Fridlund}.\\
\indent This mission, based on the implementation of nulling interferometry
\cite{Bracewell}, is devoted to the search and characterization of terrestrial exoplanets in the
mid-infrared range thanks to its combined high contrast 
and high spatial resolution (long-baseline interferometry) features \cite{Angel}.
The principle of a two aperture nulling interferometer is to recombine
artificially $\pi$ phase-shifted wavefronts from an on-axis star in order to achieve
their mutual cancellation. The signal from an off-axis object -- like a planet
orbiting the star -- is transmitted through the instrument by tuning the
interferometer baseline so that a total phase shift of 2$\pi$ is achieved.
A key requirement of nulling interferometry is 
equal wavefronts. Reducing the wavefront errors (e.g. pointing errors, optical defects due to polishing and coatings) can be obtained with single-mode waveguides as modal filters \cite{Mennesson}. Using single-mode
Integrated Optics (IO) components merges the functions of modal filtering and beam combination, providing a stable and
compact instrument with regards to external mechanical and thermal constraints. To date, the
implementation of modal filtering with single-mode waveguides has become the
major option for space-based nulling interferometry.\\
\indent The lack of single-mode waveguides for the mid-infrared in the photonics field has
pushed 
 advanced technological studies to develop such devices \cite{Borde,
  Wehmeier, Wallner}. In this context, we have initiated, under ESA contract,
the {\it IODA} program ({\it I}ntegrated {\it
    O}ptics for {\it DA}rwin) to investigate the feasibility of a
single-mode Integrated Optics concept operating over the full 5 --
20 $\mu$m spectral range. Our program has focused on the feasibility of an IO
beam combiner based on either conductive waveguides \cite{Labadie} or on
dielectric materials.\\
\indent In this paper, we present results obtained using selenide and
telluride materials to manufacture planar waveguides for the mid-infrared range, the first step toward further efforts for the development of channel waveguides. We show
that the modal behavior of the various structures can be experimentally
investigated and assessed with the {\it m}-lines method, known in the visible
and near-infrared range, and which we have successfully transferred  to the
mid-infrared range. To our knowledge, this paper presents the first study of
the modal properties of selenide and telluride thick films at longer
wavelengths.

\section{Investigating the properties of slab waveguides through {\it m}-lines study}

\subsection{Short theoretical analysis on fundamental parameters}\label{short_analysis}

\noindent It is well known that a planar waveguide must fulfill the total reflection
condition at the interface between film--substrate and film--superstrate \cite{Lee}, corresponding to $n_{\rm c}$$> $max($n_{\rm sub}$, $n_{\rm sup}$) with $n_{\rm c}$, $n_{\rm sub}$ and $n_{\rm sup}$ the refractive indices
of the film, substrate and superstrate, respectively. To propagate at least one mode, the thickness $d$ of an asymmetric planar waveguide must be greater than the cut-off thickness of the fundamental mode $d_{\rm 0}$ given, in a more general form, by:
\begin{eqnarray}
d >
d_{\rm m}&=&\f{m\pi+arctan\left(g\f{\sqrt{n_{\rm sub}^2-n_{\rm sup}^2}}{\sqrt{n_{\rm c}^2-n_{\rm sub}^2}}\right)}{k\sqrt{n_{\rm c}^2-n_{\rm sub}^2}}\label{dm}
\end{eqnarray}

\noindent where $k$=$2\pi/\lambda$ is the wave vector, $m$ the mode order starting from 0, and
$g$ a polarization factor with $g$=1 for TE polarization and $g$=$n^{2}_{\rm c}/n^{2}_{\rm sub}$ for TM polarization. Thus, the quantity $d_{\rm m}$ increases with the operating wavelength and also depends on the dispersion of the material. Here, we briefly evaluate from a simple theoretical approach the possibility to benefit of a single-mode planar waveguide over the full Darwin range (5 -- 20 $\mu$m) using As$_{2}$Se$_{3}$ glass as the layer and As$_{2}$S$_{3}$ glass as the substrate. The superstrate is air. We assume that these materials are transmissive over the entire range of interest, neglecting possible absorption lines. We also consider only the case of TE polarization with $g$=1. We derive from the refractive indices of these bulk materials \cite{Palik} a simple fit of the dispersion laws which are given, to a first approximation, by $n_{\rm c}(\lambda)$ = -0.00605$\times$$\lambda$+2.85091 and $n_{\rm sub}(\lambda)$=-0.0095$\times$$\lambda$+2.47128 for As$_{2}$Se$_{3}$ and As$_{2}$S$_{3}$ respectively, with $\lambda$ given in microns. The adopted models are justified by the fact that the refractive indices of these materials decrease almost linearly in the 5 -- 20 $\mu$m range. Under those assumptions, we compute the cut-off thicknesses $d_{\rm 0}$ and $d_{\rm 1}$   for modes $m$=0 and $m$=1 and for 5 $\mu$m$<$$\lambda$$<$20 $\mu$m. The acceptable slab thickness $d$ for single-mode behavior over our spectral band must comply with $d_{\rm 0}$(20$\mu$m)$<$$d$$<$$d_{\rm 1}$(5$\mu$m). However, since we found $d_{\rm 0}$(20$\mu$m)=2.903 $\mu$m and $d_{\rm 1}$(5$\mu$m)=2.577 $\mu$m, the previous condition cannot be fulfilled. The wavelength $\lambda_{\rm 1}$ for which $d_{\rm 0}$($\lambda_{\rm 1}$)$\le$$d_{\rm 1}$(5$\mu$m) is found to be $\lambda_{\rm 1}$=17.2 $\mu$m. Thus, it theoretically exists a film thickness, e.g. $d$=2.54 $\mu$m, for which a planar waveguide following these dispersion laws propagates the fundamental mode only over a 5 -- 17 $\mu$m band, i.e. almost the full expected band.\\
\indent We underline that this modal analysis is based only on the study of the opto-geometrical parameters of the waveguide and is thus sensitive to the numerical fit. 
To better constrain the trade-off between fundamental mode confinement and higher-mode filtering properties (which requires to be far from the cut-off), a more detailed electromagnetic approach through Beam Propagation Method (BPM) simulations will be required. However, our results show that a good compromise could reasonably be reached by splitting the 5 -- 20 $\mu$m band into no more than two sub-bands, with two dedicated waveguide designs, and on the condition that a sufficient index difference between the film and the substrate is ensured.\\

\subsection{Thick film deposition}\label{objectives}

\begin{figure}[t]
\centering
\includegraphics[width=7.9cm]{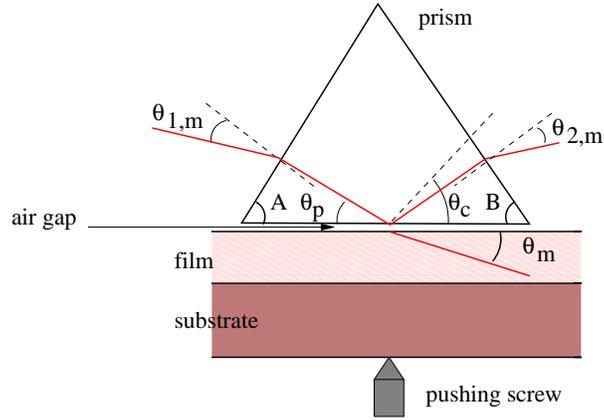}
\caption{Typical prism coupler setup. When we are in the condition of total
  reflection at the base of the prism ($\theta_{\rm p}$ smaller than critical
  angle $\theta_{\rm c}$), the modes of the slab waveguide can be excited through
  the evanescent field existing in the air gap. The air gap thickness is
  controlled with a pushing screw at the back of the
  waveguide. $A$ and $B$ are the angles of the prism. $\theta_{\rm 1,m}$ and $\theta_{\rm 2,m}$ are the algebraic angles with respect to the normal of input and output prism facets corresponding to the excitation of mode $m$. $\theta_{\rm m}$ is the angle of the propagating ray corresponding to mode $m$.}\label{principle}
\end{figure}

Extending the IO solution to the mid-IR requires infrared
materials with suitable features in this spectral domain. Chalcogenide glasses
are interesting candidates, due to their good transparency in the mid-infrared,
their high refractive index, and the relatively ease of preparation in films
form. We have selected As$_2$Se$_3$, Te$_2$As$_3$Se$_5$ and TeAs$_4$Se$_5$
chalcogenide compositions to undergo thick film deposition on an As$_2$S$_3$ substrate. Glasses containing elements like sulphur (S), selenium (Se) and tellurium (Te) transmit at longer wavelengths: sulphide glasses (i.e. As$_2$S$_3$) transmit up to 12 $\mu$m, selenide glasses (i.e. As$_2$Se$_3$) transmit up to 16 $\mu$m, and telluride glasses transmit up to 20 $\mu$m \cite{Kokorina}, with an average
transmission of about 35\% for the Te glass. Note that this transmission value includes the Fresnel losses at the interfaces due to the high refractive index of the glass. 
The deposition process used is thermal evaporation, which reduces the column-like effects in
the structure and provides dense films with higher refractive index with respect to RF-sputtering techniques.
The selenide 
films were deposited from As$_2$Se$_3$ commercial bulk glass, 
while the telluride films used bulk glass that was synthesized in the laboratory from commercial components like Tellurium (Te), Arsenic (As) and Selenide (Se).\\
\indent The strategy adopted to develop the chalcogenide planar waveguides is the
following. For both selenide and telluride compositions, we have selected
As$_2$S$_3$ chalcogenide commercial glass as the substrate -- the superstrate
is air --, which presents the lowest refractive index among the
different glasses available ($n$=2.46 at $\lambda$=1.2 $\mu$m and $n$=2.38 at
$\lambda$=10 $\mu$m). As underlined in Subsection~\ref{short_analysis}, the challenge consisted in obtaining the appropriate index difference $\Delta n=n_{\rm c}-n_{\rm sub}$ and thickness $d$ in
order to propagate at least one mode. Eq.~(\ref{dm}) shows that, for a given
$\Delta n$, $d_{\rm m}$ is proportional to $\lambda$. Thus, at longer
wavelengths the target thickness must be significantly higher than the usual
thicknesses ($\approx$ 0.5 $\mu$m) deposited for near-IR applications. Because
of the novelty of the manufactured samples, an unknown is put on the
achievable index difference and thickness. Consequently, we have experimentally
investigated the modal behavior of the sample, the refractive index and the
thickness of the films through the mid-infrared {\it m}-lines experiment
developed at LAOG (Laboratoire d'Astrophysique de Grenoble).

\subsection{Experimental technique}\label{exp_mlines}

\begin{figure}[t!]
\centering
\includegraphics[width=12cm]{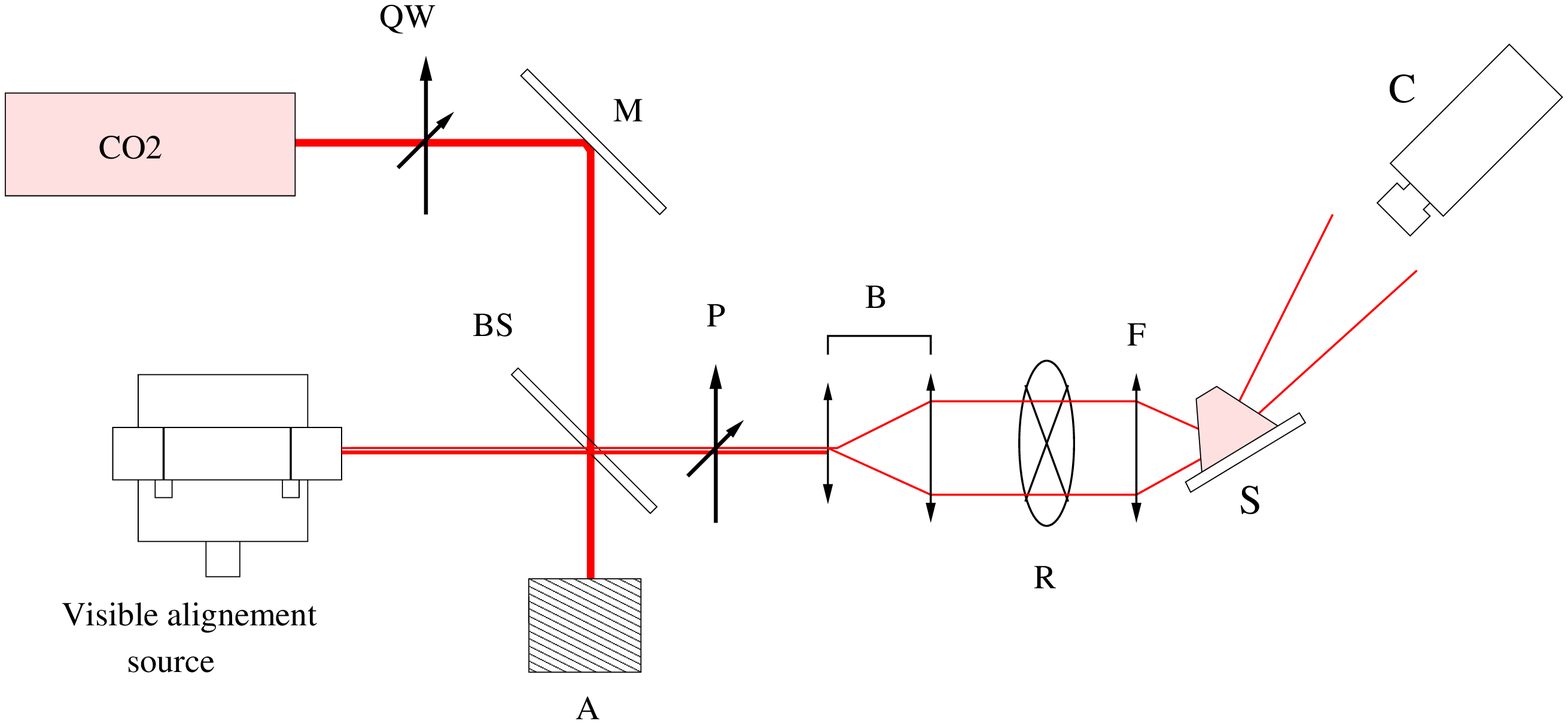}
\caption{Experimental arrangement of the mid infrared {\it m}-lines
  experiment: {\it QW}, quarter-wave plate; {\it M}, mirror; {\it BS}, beam-splitter; {\it A}, absorbing refractory brick; {\it P}, polarizer; {\it B}, beam-expander; {\it R}, reticle; {\it F}, focusing optics; {\it S}, system prism -- film; {\it C}, infrared camera.}\label{setup}
\end{figure}

\begin{figure}[t!]
\begin{minipage}{\textwidth}
\centering
\includegraphics[width=0.211\vsize]{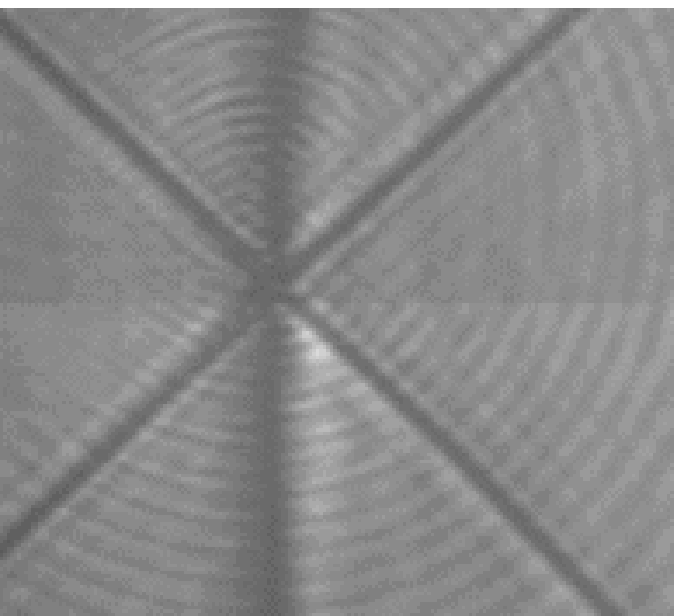}
\hspace{1cm}
\includegraphics[width=0.203\vsize]{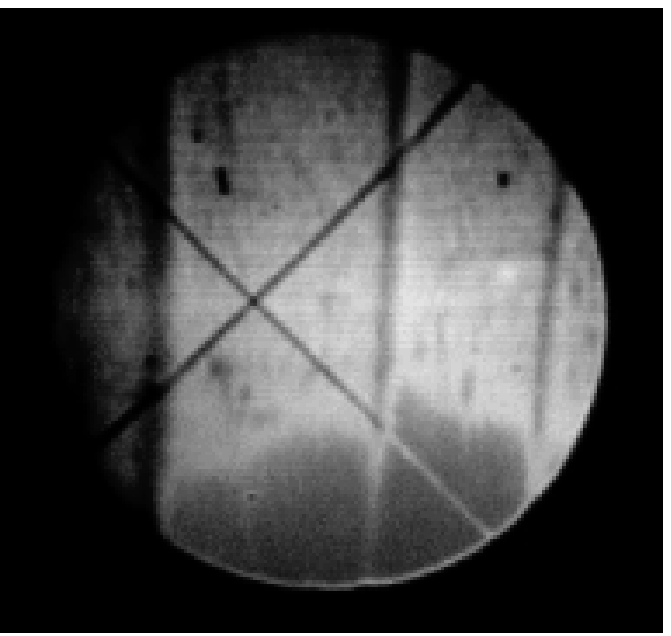}
\end{minipage}
\caption{Output of the {\it m}-lines experiment in the mid-IR (left) and in
  the near-IR (right). The image of the reticule is used to point the black
  lines by turning the prism stage. The mid-IR pattern presents a circular
  diffraction pattern due to the high coherence of the laser source.}\label{mlines-output}
\end{figure}

The {\it m}-lines method, based on prism coupling theory, was
first proposed by Tien {\it et al}. \cite{Tien} and is commonly
used as a characterization method for telecommunication components
in the visible and near-infrared range. In this study, we have
implemented this method at $\lambda$=10.6 $\mu$m. 
In this technique, a prism is brought in
close contact to the surface of a planar waveguide. Using a
pushing screw placed under the waveguide (see Fig.~\ref{principle}), a small and localized
optical contact is created between the prism base and the film.
This provides a thin air gap that should optimally be a quarter of
the operating wavelength to maximize the coupling effect.
When a ray reaches the base of the prism with an incident angle
smaller than the critical angle defined by the prism index, it
becomes totally reflected. However, the overlap of the evanescent
fields in the air gap as predicted by electromagnetic theory \cite{Tien2} implies that light can be coupled into the waveguide
similarly to a tunneling effect. The existing modes of the
waveguide will be excited if the phase matching conditions given
by Eq.~(\ref{phase-match}) are fulfilled

\begin{eqnarray}
n_{\rm p}\cos(\theta_{\rm p})&=&n_{\rm c}\cos(\theta_{\rm m})\label{phase-match}
\end{eqnarray}


\noindent where $n_{\rm p}$ and $n_{\rm c}$ are respectively the prism and the thick film indices, and $\theta_{\rm p}$ and
$\theta_{\rm m}$ are the angular directions of the rays as illustrated in
Fig.~\ref{principle}. Exciting the low-order modes of the structure requires a prism with higher index than the slab waveguide index. Under these
conditions, and for the incident rays that fulfill the phase conditions, the
coupling phenomenon will lead to the observation of "black lines" at discrete
positions $\theta_{\rm 2,m}$ in the reflected beam pattern. These lines are
characteristic of the different modes of the slab waveguide. From the measurement of the angles $\theta_{\rm
  2,m}$, the effective index $n_{\rm eff,m}$ of mode $m$ is computed through:

\begin{eqnarray}
n_{\rm eff,m}&=&n_{\rm p}\sin\left(\arcsin\left(\f{\sin(\theta_{\rm 2,m})}{n_{\rm p}}\right)+B\right)\label{neff}
\end{eqnarray}

\noindent where $B$ is one of the prism angles. 
The computation of the thick film index $n_{\rm c}$ and its thickness $d$ is obtained by
implementing a non-linear least squares method (see Subsection~\ref{data_process}).\\ 
\indent \rm The experimental setup to measure the mode indices
at 10.6 $\mu m$ is shown in Fig.~\ref{setup}. The infrared source is a CO$_{\rm
  2}$ laser. To probe either TE or TM modes, the quarter-wave plate $QW$ modifies the intrinsic linear polarization of the source into an elliptical one, then the grid polarizer $P$ selects the desired polarization. The Zinc Selenide (ZnSe) optics $B$ and $F$ produce a 
converging beam which is focused on the optical contact. This beam includes the angular directions that fulfill Eq.~(\ref{phase-match}), simplifying the search for mode lines. The reticule is used to align the normal of
the input face of the prism with the optical axis of the bench in order to
solve Eq.~(\ref{neff}). The black lines in the reflected beam are observed with a
mid-infrared camera (see Fig.~\ref{mlines-output}).\\
\indent For the experiment at 10.6 $\mu$m, we used a high-index germanium prism with
an angle $B$=30$^{\circ}$$\pm$0.01$^{\circ}$ and a refractive index $n_{\rm
  p}$=4.0$\pm$0.001. In one case, the {\it m}-lines experiment was run at shorter
wavelengths ($\lambda$=1.196 $\mu$m) and for which we used a silicon prism with an
angle $B$=44.8416$^{\circ}$$\pm$5$\times$10$^{-4}$ and a refractive index $n_{\rm
  p}$=3.5193$\pm$10$^{-4}$. In our system, the position of the black line is
measured with an accuracy of $\pm$0.01$^{\circ}$.




\subsection{Data processing}\label{data_process}

In this paper, we have assumed that the tested samples are asymmetric slab
waveguides (i.e. Air/As$_{\rm 2}$Se$_{\rm 3}$/As$_{\rm 2}$S$_{\rm 3}$) with a step index distribution. To obtain a correct numbering of
the modes, we used the property of linearity between the squared spectrum of
mode indices $n^2_{\rm eff}(m)$ and the mode numbers set ($m$+1)$^2$, as formalized by Lee {\it et al.}\cite{Lee88}{}.
Thus, when more than three experimental mode indices are available, their
numbering is derived from the computation of the linear regression coefficient
$\rho$ 
of the experimental curve [$n^2_{\rm eff}(m)$,(m+\rm 1)$^2$] and
compared to unity (see Tables~\ref{As-1IR},\ref{As-1IR-10um}). In this paper, we assume that mode numbering starts with $m$=0. Successively, we compute $n_{\rm c}$ and $d$ by numerically
solving \cite{White} the waveguide dispersion equation for a step index
distribution planar waveguide given by Eq.~(\ref{disp}) \cite{Lee}:

\begin{eqnarray}
kd\sqrt{n_{\rm c}^{2}-n_{\rm eff,m}^{2}}-\arctan\left(g\f{\sqrt{n_{\rm
      eff,m}^{2}-n_{\rm sub}^{2}}}{\sqrt{n_{\rm c}^{2}-n_{\rm
      eff,m}^{2}}}\right)
-\arctan\left(g\f{\sqrt{n_{\rm
      eff,m}^{2}-n_{\rm sup}^{2}}}{\sqrt{n_{\rm c}^{2}-n_{\rm
      eff,m}^{2}}}\right)
=m\pi \label{disp}
\end{eqnarray} \\

\indent To better constrain the result given by the linear regression
coefficient method, we use the derived quantities $n_{\rm c}$ and $d$ to
compute the theoretical mode indices for different mode numbering by
analytically solving Eq.~(\ref{disp}). The correct mode numbering is then
obtained through a best fit between the theoretical and experimental distributions of indices.

\section{Results and discussion}

\subsection{Selenide components}

\noindent 
The samples, with a target thickness around 10
  $\mu$m, were characterized in the near-IR ($\lambda$=1.2 $\mu$m) and in the
  mid-IR ($\lambda$=10.6 $\mu$m) with the {\it m}-lines setup presented
  in Subsection~\ref{exp_mlines} \rm.
Two samples were produced and respectively referenced as {\bf As-1} and {\bf
  As-2}. 
Sample {\bf As-1} was characterized both at $\lambda$=1.2 $\mu$m and at
  $\lambda$=10.6 $\mu$m, while sample {\bf As-2} was characterized in
  the mid-IR only, Including the near-IR characterization is advantageous, since a multimode behavior is expected from the moment the film
  thickness is greater than 1 $\mu$m. Thus, this increases the number of mode
  lines that can be detected and measured.

\begin{table}[t]
\centering
\begin{tabular}{|c|c|c|c|c|c|} \hline
Mode & Measured & Angle      & Experimental     & Theoretical & Diff.  \\
     & angle    & separation & n$_{\rm eff}$    & n$_{\rm eff}$   &  \\
\hline
 0 & a                 & a     & N/A    & 2.7729 & N/A     \\ \hline
 1 & a                 & a     & N/A    & 2.7719 & N/A     \\ \hline
 2 & +25.65$^{\circ}$  & a     & 2.7697 & 2.7702 & 0.0005  \\ \hline
 3 & +25.50$^{\circ}$  & +0.15 & 2.7682 & 2.7679 & 0.0003  \\ \hline
 4 & +25.21$^{\circ}$  & +0.29 & 2.7654 & 2.7649 & 0.0005  \\ \hline
 5 & +24.81$^{\circ}$  & +0.4  & 2.7614 & 2.7613 & 0.0001  \\ \hline
 6 & +24.39$^{\circ}$  & +0.42 & 2.7573 & 2.7570 & 0.0003  \\ \hline
 7 & +23.79$^{\circ}$  & +0.6  & 2.7513 & 2.7519 & 0.0006  \\ \hline
 8 & +23.25$^{\circ}$  & +0.54 & 2.7459 & 2.7462 & 0.0003  \\ \hline
 9 & +22.67$^{\circ}$  & +0.58 & 2.7400 & 2.7399 & 0.0001  \\ \hline
\multicolumn{2}{|l}{Refractive index $n_{\rm c}$
}&\multicolumn{4}{l|}{2.7732 $\pm$1.3$\times$10$^{-4}$}
\\ \hline
\multicolumn{2}{|l}{Thickness $d$}&\multicolumn{4}{l|}{13.74 $\pm$ 0.05
     $\mu$m} \\ \hline
\end{tabular}
\caption{TE mode indices obtained experimentally on sample {\bf As-1} at
  $\lambda$=1.196 $\mu$m with the {\it m}-lines setup. The error on the
  measured angle is $\pm$0.01$^{\circ}$. The error on experimental n$_{\rm
  eff}$ is $\pm$1.3$\times$10$^{-4}$. We found a maximum coefficient
  $\rho$=-0.99923 for the presented mode numbering. The silicon prism has
  n$_{\rm p}$=3.5193 and B=44.8416$^{\circ}$. The substrate refractive index
  at $\lambda$=1.196 $\mu$m is 2.47\cite{Palik}. Note that the two first modes
  could not be experimentally observed with the silicon
  prism and are labelled as "$a$" for "absent". The sign "N/A" means "not applicable".}\label{As-1IR}\end{table}
\vspace{0.25cm}

\indent Table~\ref{As-1IR} reports the TE mode indices and the
derived parameters for {\bf As-1} at $\lambda$= 1.196 $\mu$m. Sixteen mode lines
were experimentally observed with the sample {\bf As-1}, but only the first
eight modes are presented here. 
Note that measurements of TM modes were not carried out because the two optical components $QW$ and $P$ in Fig.~\ref{setup} were not available at the time of the corresponding tests. Thus, only the intrinsic linear polarization of the source, corresponding to TE modes, was used for those two samples.
The derived quantities are $n_{\rm c}$=2.7732 $\pm$ 1.3$\times$10$^{-4}$ and $d$=13.74 $\pm$ 0.05 $\mu$m. 
A similar characterization was made at $\lambda$=1.55 $\mu$m, in order to
observe a chromatic change in refractive index. This led to values 
$n_{\rm c}$(1.55$\mu$m)=2.7407$\pm$1.3$\times$10$^{-4}$ and $d$=13.16 $\pm$ 0.03 $\mu$m. We observe a change in refractive
index of approximately 0.04 as well as a small variation of the film
thickness. Note that the {\it m}-lines experiment provides a measurement of
the pair of quantities ($n_{\rm c}$,$d$) for the film area where the optical contact is
obtained. When 
shifting from $\lambda$=1.2 $\mu$m to $\lambda$=1.55 $\mu$m, we also moved the prism on the film, which has an effect on the measured film thickness.\\
\indent We performed {\it M}-lines characterization in the mid-infrared at
$\lambda$=10.6 $\mu$m for the two samples {\bf As-1} and {\bf As-2}. At longer
wavelengths, the germanium prism described in Subsection~\ref{exp_mlines} \rm was
used. Table~\ref{As-1IR-10um} shows the experimental results obtained with the two
samples. The value of the substrate index, $n_{\rm s}$=2.380, has been taken
from the literature \cite{Palik}. 



\begin{table}[t!]
\centering
\begin{tabular}{|c|c|c|c|c|c|} \hline
\multicolumn{3}{|l}{sample {\bf As-1}}&
\multicolumn{3}{l|}{} \\ \hline
Mode & Measured & Angle      & Experimental     & Theoretical & Diff.  \\
     & angle    & separation & n$_{\rm eff}$    & n$_{\rm eff}$   &  \\
\hline
 0 & +55.04$^{\circ}$  & N/A   & 2.667  &  2.6666  & 0.0004 \\ \hline
 1 & +46.89$^{\circ}$  & 8.15  & 2.599  &  2.5992  & 0.0002 \\ \hline
 2 & +36.09$^{\circ}$  & 10.8  & 2.488  &  2.4868 & 0.0012    \\ \hline
 \multicolumn{2}{|l}{Refractive index $n_{\rm c}$
}&\multicolumn{4}{l|}{2.689 $\pm$ 0.001}
\\ 
\multicolumn{2}{|l}{Thickness $d$}&\multicolumn{4}{l|}{13.27 $\pm$ 0.03 $\mu$m}
\\ \hline \hline
\multicolumn{3}{|l}{sample {\bf As-2}} & \multicolumn{3}{l|}{} \\ \hline
0 & +48.58$^{\circ}$  & N/A   & 2.614  &  2.6138  & 0.0002  \\ \hline
1 & +42.61$^{\circ}$  & 5.97  & 2.557  &  2.556  & 0.001  \\ \hline
2 & +33.61$^{\circ}$  & 9.00  & 2.4601  &  2.4606  & 0.0005  \\ \hline
 \multicolumn{2}{|l}{Refractive index $n_{\rm c}$} &\multicolumn{4}{l|}{2.633
   $\pm$ 0.001} \\
\multicolumn{2}{|l}{Thickness $d$}&\multicolumn{4}{l|}{14.49 $\pm$ 0.03 $\mu$m} \\ \hline
\end{tabular}
\caption{TE mode indices obtained experimentally on samples {\bf As-1} and
  {\bf As-2} at $\lambda$=10.6 $\mu$m. The error on measured angle is
  $\pm$0.01$^{\circ}$. The accuracies on prism index and angle lead to an
  error on experimental n$_{\rm eff}$ of $\pm$10$^{-3}$. The
  maximum coefficient is $\rho$=-0.99982 for the given mode numbering. The
  prism has n$_{\rm p}$=4.00$\pm$0.001 and B=30.0$\pm$0.01$^{\circ}$. The
  substrate refractive index at $\lambda$=10.6 $\mu$m is 2.38. The sign "N/A" means "not applicable".}\label{As-1IR-10um}\end{table}


\indent  For the two samples, three mode lines have been observed with our
 setup. 
 The derived theoretical mode indices are very close to the experimental
 ones. The dispersion of the mode indices is low, which implies a good
 homogeneity of the index distribution. The derived refractive indices are
 $n_{\rm c}$=2.689 $\pm$ 0.001 for {\bf As-1} and $n_{\rm c}$=2.633 $\pm$ 0.001 for {\bf As-2}, which shows that the densification of the layer was
 lower during the second run. The sample thickness is approximately 13 $\mu$m
 for {\bf As-1} and is consistent with the value obtained at shorter
 wavelengths. The thickness is slightly higher for {\bf As-2} with a value of
 approximately 14.5 $\mu$m.

\indent We measure mid-IR refractive indices significantly lower than in the
near-IR. This is consistent with the dispersion law of the material, which
predicts a decrease in the refractive index with increasing wavelength.
We also note that the achieved thicknesses are greater than the targeted ones, which shows the deposition process needs to be better stabilized. Finally, a multimode behavior of the planar structures is clearly identified
in the near and mid-IR ranges.

\subsection{Telluride components}


\begin{table}[t!]
\centering
\begin{tabular}{|c|c|c|c|c|c|} \hline
\multicolumn{3}{|l}{sample {\bf TAS-1} (TM polarization)} &
\multicolumn{3}{l|}{} \\ \hline
Mode & Measured & Angle      & Observed       & Theoretical   & Diff .\\
     & angle    & separation & n$_{\rm eff}$  & n$_{\rm eff}$ &       \\ \hline
 0   & +55.83$^{\circ}$   & N/A          & 2.673          & 2.672         & 0.001   \\ \hline
 1   & +43.56$^{\circ}$   & +12.27     & 2.567          & 2.561         & 0.006   \\ \hline
 2   & +28.13$^{\circ}$   & +15.43     & 2.394          & 2.395         & 0.001   \\ \hline
\multicolumn{3}{|l}{Refractive index $n_{\rm c}$} &
\multicolumn{3}{l|}{2.709 $\pm$ 0.001} \\ 
\multicolumn{3}{|l}{Thickness $d$}& 
\multicolumn{3}{l|}{10.72 $\pm$ 0.03 $\mu$m} \\ \hline \hline
\multicolumn{3}{|l}{sample {\bf TAS-1} (TE polarization)} &
\multicolumn{3}{l|}{} \\ \hline
Mode & Measured & Angle       & Observed      & Theoretical   & Diff.  \\
     & angle    & separation  & n$_{\rm eff}$ & n$_{\rm eff}$ &        \\ \hline
 0   & a                  &   N/A       & a             & 2.686        & N/A      \\ \hline
 1   & +45.26$^{\circ}$   &   N/A       & 2.583        & 2.5887        & 0.0057 \\ \hline
 2   & +30.13$^{\circ}$   &   +15.13    & 2.419        & 2.4298        & 0.0108 \\ \hline
\multicolumn{3}{|l}{Refractive index $n_{\rm c}$} &
\multicolumn{3}{l|}{2.718 $\pm$ 0.001} \\ 
\multicolumn{3}{|l}{Thickness $d$}& 
\multicolumn{3}{l|}{10.69 $\pm$ 0.03 $\mu$m} \\ \hline 
\end{tabular}
\caption{TM (up) and TE (down) mode indices obtained experimentally on sample
  {\bf TAS-1} at $\lambda$=10.6 $\mu$m. The error on measured angle is
  $\pm$0.01$^{\circ}$. The accuracies on prism index and angle lead to an
  error on experimental n$_{\rm eff}$ of $\pm$10$^{-3}$. The same germanium
  prism as in Table~\ref{As-1IR-10um} is used. The sign "$a$" means "absent" and "N/A" means "not applicable".}\label{TAS-10um}
\end{table}


\noindent 
In this approach, the thick films were deposited by evaporation on a As$_{\rm 2}$S$_{\rm 3}$ substrate
from 
Te$_{\rm 2}$As$_{\rm 3}$Se$_{\rm 5}$ and TeAs$_{\rm 4}$Se$_{\rm 5}$ bulk
glasses made by LPMC (Laboratoire de Physico-chimie de la Mati\`ere
  Condens\'ee). Two samples referenced {\bf TAS-1} (TeAs$_{\rm 4}$Se$_{\rm
  5}$/As$_{\rm 2}$S$_{\rm 3}$) and {\bf TAS-2} (Te$_{\rm 2}$As$_{\rm
  3}$Se$_{\rm 5}$/As$_{\rm 2}$S$_{\rm 3}$) were produced by thermal
evaporation and characterized for TE and TM polarizations at
$\lambda$=10.6 $\mu$m. The same germanium prism was used to achieve light
coupling. The experimental measurements on {\bf TAS-1} are reported in
Table~\ref{TAS-10um}. For {\bf TAS-2}, we present the derived refractive
index and thickness only in Table~\ref{TAS-2-10um}. Mode lines have been observed
for the two samples. The numerical analysis confirmed that the dark lines
correspond to propagating modes. In this case as well, more than one line were
observed.\\
\indent For sample {\bf TAS-1}, the fundamental mode in TE polarization could not be
observed, oppositely to the TM case. For each mode, the measurement of the line
angular position for the two polarizations was done successively. In this way,
we could experimentally observe that there was a slight angular shift for the same mode between the two polarization. The {\it m}-lines measurement was done
for TE and TM polarizations at the same point of the film. The results show a
small difference between TM and TE index $n_{\rm c}$, which supposes an
hypothesis of birefringence of the film. For this first telluride component,
an approximate thickness of the film of about 10 $\mu$m has been achieved.\\
\indent The sample {\bf TAS-2} was characterized with a similar
procedure. Three mode lines have been experimentally observed but only in TE
polarization. The corresponding TM case could not be observed despite our
efforts in optimizing the coupling efficiency. For {\bf TAS-2}, we therefore present the derived refractive index and thickness only in Table~\ref{TAS-2-10um}.
\newpage

\begin{table}[t!]
\centering
\begin{tabular}{|c|c|c|c|c|c|} \hline
\multicolumn{3}{|l}{sample {\bf TAS-2} (TE polarization)} &
\multicolumn{3}{l|}{} \\ \hline
Mode & Measured & Angle      & Observed       & Theoretical   & Diff .\\
     & angle    & separation & n$_{\rm eff}$  & n$_{\rm eff}$ &       \\ \hline
 0   & +73.59$^{\circ}$   & N/A        & 2.772          & 2.776         & 0.004   \\ \hline
 1   & +52.45$^{\circ}$   & +21.14     & 2.647          & 2.640         & 0.007   \\ \hline
 2   & +30.00$^{\circ}$   & +22.45     & 2.417          & 2.419         & 0.002   \\ \hline
\multicolumn{3}{|l}{Refractive index $n_{\rm c}$} &
\multicolumn{3}{l|}{2.821 $\pm$ 0.001} \\ 
\multicolumn{3}{|l}{Thickness $d$}& 
\multicolumn{3}{l|}{8.79 $\pm$ 0.03 $\mu$m} \\ \hline 
\end{tabular}
\caption{TE mode indices obtained on sample {\bf TAS-2} at $\lambda$=10.6
  $\mu$m. The error on measured angle is $\pm$0.01$^{\circ}$ and the accuracies on prism index and angle lead to an
  error on experimental n$_{\rm eff}$ of $\pm$10$^{-3}$.}\label{TAS-2-10um}
\end{table}


\indent Compared to sample TAS-1, TAS-2 presents a higher refractive index
at $\lambda$=10 $\mu$m. The possibility to obtain a high refractive index is
an important result if a ``full-telluride'' (film and substrate) is foreseen. Using a telluride
glass as a substrate will indeed raise the question of the achievable
index difference between the film and the substrate, since telluride
bulks are expected to present higher index compared to selenide bulks.

\subsection{Impact of the experimental results on the requirements for single-mode slab waveguides}

\noindent These experimental results show that the refractive index obtained with As$_{\rm 2}$Se$_{\rm 3}$ under thick film form is lower (by approximately 0.15) than in bulk form \cite{Palik}. 
Since the index difference between the film and the substrate is an important parameter, we have now a less favorable case compared to the analysis of Subsection~\ref{short_analysis}. Deriving the new model $n_{\rm c}(\lambda)$ = -0.0127$\times$$\lambda$+2.78532 from the data of Table~\ref{As-1IR} and~\ref{As-1IR-10um}, we find that we would need two As$_{\rm 2}$Se$_{\rm 3}$/As$_{\rm 2}$S$_{\rm 3}$ slab waveguides to cover the 5 -- 20 $\mu$m band: the first one with a thickness $d$=3.11 $\mu$m for the sub-band 5 -- 12.5 $\mu$m, and the second one with a thickness $d$=7.0 $\mu$m for the sub-band 12.5 -- 20 $\mu$m.\\
\indent Extending the first single-mode sub-band can be achieved with the telluride glass Te$_{\rm 2}$As$_{\rm 3}$Se$_{\rm 5}$ as the guiding layer, since the experimental results of Table~\ref{TAS-2-10um} show that the achievable refractive index is higher than for As$_{\rm 2}$Se$_{\rm 3}$.\\
\indent However, 
the As$_{\rm 2}$S$_{\rm 3}$ substrate transparency remains an issue -- it does not extend up to 20 $\mu$m. Therefore, covering the whole 5 -- 20 $\mu$m band would require an examination of a 20 $\mu$m transparent telluride substrate, whose refractive index is sufficiently low to ensure a $\Delta$n of about 0.42.

\section{Conclusions}


\noindent This paper has discussed the optical characterization of
chalcogenide-glass-based thick films manufactured for the purpose of future mid-IR
integrated optics development. We have fully developed an {\it m}-lines setup
operating at $\lambda$=10.6 $\mu$m based on prism coupling theory. This setup
was used to characterize the modal behavior of the first mid-infrared planar
waveguides by observing mode lines. In this study, two different compositions were successfully tested at 10 $\mu$m with the first observation of three mode lines for TE and TM polarizations. A simple numerical analysis based on a
best fit between the experimental and theoretical mode indices distributions has permitted us to derive the refractive index and thickness of the different samples. Our results have underlined the possibility of producing full-selenide
multimode planar waveguides with a sufficient film thickness. The near and mid-IR results are consistent, since similar thicknesses are obtained in the two spectral ranges. Extending the transmission range of the components to 20 $\mu$m
may be possible with the telluride solution.\\
\indent Following these results, a complementary technological study has been conducted to
investigate the feasibility of a first full-telluride single-mode planar
waveguide. This study has produced positive results. As a consequence, the question of
etching of a channel waveguide to be used as a modal filter for nulling
interferometry has been investigated straightforward and the results recently
obtained will be part of a forthcoming paper.

\section*{Acknowledgments}

\noindent This work was funded by European Space Agency contract
16847/02/NL/SFe, French Space Agency (CNES) and Alcatel-Alenia Space. The
authors are grateful to Amal Chabli and Tom Herbst for useful discussions.

\end{document}